\documentclass[10pt,twocolumn]{revtex4-2} 
\usepackage[dvips]{color}
\usepackage{epsfig}
\usepackage{amssymb}
\usepackage{latexsym}
\usepackage{graphicx}
\begin{document}
\title{Analysis of Phonetic Soliton Propagation in Neutral Weyl Fermion-sea}

\author{ Sadataka Furui}\affiliation{Faculty of Science and Engineering, Teikyo University, Utsunomiya, 320 Japan }
 \email{furui@umb.teikyo-u.ac.jp}
\author
{ Serge Dos Santos}\affiliation{INSA Centre\, Val de Loire,  Blois,
Inserm U1253, Universit\'e de Tours, Imagerie et Cerveau, imaging and brain : iBrain, France }
\email{serge.dossantos@insa-cvl.fr} 

\date{\today }

\begin{abstract}
We propose application of Machine Learning (ML) and Neural Network (NN) technique for the analysis of ultrasonic Time Revesal based Nonlinear Elastic Wave Spectroscopy (TR-NEWS).   

In order to acquire topological features, we adopt the $(2+1)D$ lattice simulation with fixed point (FP) actions. We consider 7 A type loops which sit on $2D$ spacial plane spanned by $e_1, e_2$ and 13 B type loops which contain links parallel to $e_1\wedge e_2$ and to -$e_1\wedge e_2$.
  
We consider propagation of bosonic phonons in Fermi-sea of neutral Weyl spinors which are described by Clifford algebra. Configurations in momentum space is transformed to real position space via Clifford Fourier Transform.

We consider A-type without hysteresis effects and B-type with hysteresis effects, and via ML or NN technique
search optimal weight of 7 A-type FP actions and 13 B-type FP actions using the Monte-Carlo method.

\end{abstract}
\maketitle

\section{Introduction}
In non-destructive-testing (NDT) the time reversal (TR) based nonlinear elastic wave spectroscopy (TR-NEWS) was successful. Use of time reversal mirrors (TRM) for enhancing signal to noise ratios of ultrasonic phonetic waves scattered in materials was proposed by Fink\cite{Fink97} and the technique of focussing, decomosition of time reversal operator (DORT) was developed\cite{RBCBF06}.

In the TR-NEWS, transducers which emit phonons and their TR phonons and receivers are placed on sides of a 2 dimensional plane\cite{GDSBMC08,DSVP09}. Propagation of nonlinear phonetic waves in materials is recently reviewed in \cite{PM22}. 

Haldane\cite{Haldane04} revealed the spin quantum Hall effect (QHE) in space 2 dimensional TR violating electronic system can be explained through Chern index of algebraic topology. 

Kane and Mele\cite{KM05} studied the QHE in TR symmetric systems and showed that $Z_2$ topological order characterize presence of ordered phase on the Fermi surface.

Fidkowski and Kitaev\cite{FK11} discussed that TR invariant Majorana fermion spin 1 system has a trivial and Haldane phase, and the breaking of $Z$ symmetry to $Z_2$ symmetry can be interpreted by real (R) edge spin states and quaternion (H) edge spin states. 

 Ryu et al.\cite{RML12} studied electromagnetic and gravitational responses and anomalies in topological insurators and superconductors by extending the theory of \cite{KM05}. We want to apply these theories to TR symmetric propagation of phonons in fermionic media. 

Since the media that a phonon propagates is not charged, we construct a model of Weyl fermions represented by quaternions sitting on $(2+1)D$ lattices\cite{SF21b,SF21c}. Instead of electric flow that propagates in the background of gravity, we consider the energy flow described by the DORT method.

\section{ESAM and DORT method}
The nonlinear waves are expressed as a linear combination of $x(t), x^2(t)$ and $x^3(t)$, and symmetry of the point group of $x(t), x^2(t), x^3(t)$ leads to $O(3)$ transformed bases $x_E(t)=x(t)=y_E$, $x_\epsilon(t)= x(t)e^{\sqrt{-1}\pi/3}=y_{B1}$, $x_{\epsilon^*}=x(t)e^{-\sqrt{-1}\pi/3}=y_{B2}$ and $-(1/2)x(t)=y_A$. The bases $y_A, y_E, y_{B1}$ and $y_{B2}$ can be identified as those of a quaternion and used in the ESAM (Excitation Symmetry Analysis). We perform Quaternion Fourier Transform (QFT) and map to the $(2+1)D$ momentum space of Clifford algebra with bases $e_1,e_2$ and $e_{12}$. We choose $e_{12}$ as a temporal basis.
 
The input signals are assumed to have cubic expansion in nonlinear waves response is written as
\[
y(t)={\mathcal N\,L}[x(t)]=N_1 x(t)+N_2 x^2(t)+N_3 x(t)^3
\]
and the intrinsic parameters $N_1, N_2$ and $N_3$ are defined by the relation
\begin{equation}
\left(\begin{array}{c}
N_1 x(t)\\
N_2 x^2(t)\\
N_3 x^3(t)\end{array}\right)=\frac{1}{3}\left(\begin{array}{cccc}
-1& -8 & 2 & 2\\
0 & 0 & 2 & 2 \\
4 & 8 & -4 & -4\end{array}\right)\left(\begin{array}{c}
y_E\\
y_A\\
y_{B1}\\
y_{B2}\end{array}\right). 
\end{equation}

For $x_E(t)=x(t)$ and $x_A(t)=-\frac{1}{2}x(t)$, the simultaneous equation
\begin{eqnarray}
x(y_E,y_{B1},y_{B2})=(3y_E+2y_{B1}+2y_{B2})/(3 N_1)\nonumber\\
x(y_E,y_{B1},y_{B2})^2=(2y_{B1}+2y_{B2})/(3 N_2)\nonumber\\
x(y_E,y_{B1},y_{B2})^3=(-4y_{B1}-4y_{B2})/(3 N_3)
\end{eqnarray}
give following solutions:
\begin{eqnarray}
&&
\frac{N_1}{N_3^{1/3}}=-\frac{3 y_E+2 y_{B1}+2y_{B2}} {6^{2/3}(y_{B1}+ y_{B2})^{1/3}}, \nonumber\\
&&\frac{N_2}{N_3^{2/3}}=\frac{Y_1}{6^{1/3} ( y_{B1}+y_{B2})^{2/3} Y_2}
\end{eqnarray}

\begin{eqnarray}
&&\frac{N_1}{N_3^{1/3}}=-\frac{3 y_E+2 y_{B1}+2y_{B2}}{ 6^{2/3}( y_{B1}+ y_{B2})^{1/3}}, \nonumber\\
&&\frac{N_2}{N_3^{2/3}}=-\frac{Y_1}{6^{1/3}( y_{B1}+y_{B2})^{2/3}Y_2}
\end{eqnarray}

\begin{eqnarray}
&&\frac{N_1}{N_3^{1/3}}=-\frac{3 y_E+2 y_{B1}+2y_{B2}}{(-6)^{2/3}( y_{B1}+ y_{B2})^{1/3}}\nonumber\\
&&\frac{N_2}{N_3^{2/3}}=-\frac{Y_1}{(-6)^{1/3} ( y_{B1}+y_{B2})^{2/3}Y_2}
\end{eqnarray}
where 
\begin{eqnarray}
&&Y_1=4 y_{B1}^3+12 y_{B1}^2 y_{B2}+12 y_{B1}y_{B2}^2+4y_{B2}^3+12 y_{B1}^2 y_E\nonumber\\
&&+24 y_{B1}y_{B2} y_E+ 12 y_{B2}^2 y_E
+9y_{B1} y_E^2+9 y_{B2}y_E^2\nonumber\\
&&Y_2=4y_{B1}^2+8y_{B1} y_{B2}+4y_{B2}^2+12y_{B1}y_E\nonumber\\
&&+12 y_{B2}y_E+9y_E^2.
\end{eqnarray}

\section{The Normalizing Flow expressed by Quaternions}
The nonlinear energy flow $f({\bf x})$ expressed by quaternions are Fourier transformed\cite{Hitzer22},
\begin{equation}
\hat f({\bf u})=\int_{R^2} e^{-{\bf i}x_1 u_1}f({\bf x})e^{-{\bf j}x_2 u_2}d^2{\bf x}
\end{equation}

  The inverse Clifford transform (ICFT) of ${\mathcal F}\{f\}\in L^1(R^2;{\mathcal G}_2)$ is
 \begin{equation}
f({\bf x})=\frac{1}{(2\pi)^2}\int_{R^2} {\mathcal F}({\bf \omega})e^{i_2 \omega}d^2{ \omega}\nonumber\\
  \end{equation}
  where  $\omega=u_1 e_1+u_2  e_2+\omega_{12}e_{12}$, ${\bf x}=x_1 e_1+x_2 e_2$,
 \begin{eqnarray}
 &&F=\{ f_1({\bf x,u}), f_2( {\bf x,u})\}, \nonumber\\
 &&f_1({\bf x, u})=2\pi e_1 x_1 u_1, f_2({\bf x, u})=2\pi e_2 x_2 u_2,
\end{eqnarray}
is supposed to be invertible. 

We let ${\bf u}$ be a $(2+1)D$ vector projected on $2D$ represented by quaternions, and transformation $\tau$ yields real vector space ${\bf x}=\tau({\bf u})$, where ${\bf u}\sim p_L({\bf u})$, where $L$ distinguishes 20 loops.
 
 The normalizing flow proposed by Papamakarios et al.\cite{PNRML21} and reviewed in \cite{Ye22} for the convolutional neural network (CNN) can be applied to the search of 7 dimensional optimal weight functions of $A$ type actions.
 
 We transform the momentum space distribution $p_L({\bf u})=\sum_k^E p^k_L({\bf u})$ into random number distribution $(0,1)^E$, where $E$ is the number of epochs.  The distribution at the epoch $k$ is related to that at the epoch $k-1$ by ${\bf u}_k=T_k({\bf u}_{k-1})$. The inverse is ${\bf u}_{k-1}=T^{-1}_k({\bf u}_{k})$.
 
 To every distribution $u_i$, we consider the Clifford Fourier transform (CFT) $x_i=\tau(u_i, {\bf h}_i)$ where ${\bf h}_i=c_i({\bf u}_{<i})$, and $\tau$ is the transformer and $c_i$ is the conditionner at the $i$th epoch\cite{Hitzer22}.  
 
The position space transformation from $\bf u$ to $\bf x$, which is the inverse CFT is defined as
\begin{equation}
 h({\bf x})= {\mathcal F}^{-1}\{{\bf x}\}({\bf u})=\frac{1}{(2\pi)^2}\int_{R^2}e^{f x_1 u_1}{\mathcal F}\{ h\}({\bf u})
 e^{g x_2 u_2}d^2{\bf u}
\end{equation}
 where $d^2{\bf u}=d u_1 d u_2$, $f,g\in {\bf H}\sim Cl(0,2), f^2=g^2=-1$, or $f,g\in Cl(2,0)$.

 In the case of $A$ type loops, we consider simple $2D$ QFT to obtain $f({\bf x})$, from ${\bf \omega}=\omega_{01}e_1+\omega_{02}e_2$ as in \cite{Hitzer22}. However, in the case of $B$ type loops, we  include the scalar term and consider the energy-time transformation.
 
  Given $x_i$ we can compute $u_i=\tau^{-1}(x_i,{\bf h}_i)$.
  
  Since $x_i$ depends only on $x_j (j<i)$, the Jacobian of the transformation $\tau$ 
\begin{equation}
  J_{\tau}({\bf u})=\left(\begin{array}{ccc}
  \frac{\partial \tau}{\partial u_1}(u_1, {\bf h}_1  
  ) & \cdots & {\bf 0}\\
  \vdots &  \ddots & \vdots\\
  {\bf L}({\bf u})&\cdots & \frac{\partial \tau}{\partial u_E}(u_E,{\bf h}_E)\end{array}\right)
\end{equation}
Corresponding to $p_L({\bf u})$ the flow 
\begin{equation}
p_F({\bf x})=p_L(T^{-1}({\bf x})) |{\rm det} J_{T^{-1}}({\bf x})|,
\end{equation}

 The $T_k$ transforms ${\bf z}_{k-1}$ to ${\bf z}_k$,  and multiple of $T_1 T_2 \cdots T_{E}$ transforms ${\bf z}_0={\bf u}$ to ${\bf z}_E={\bf x}$.
 
 The inverse flow $T^{-1}=T_1^{-1}\circ T_2^{-1}\circ\cdots\circ T_E^{-1}$ takes a collection of samples from $p_F({\bf x})$ and transform them (in a sense `normalizes' them) to a collection of samples from the density $p_L({\bf u})$. Suppose that $p_F({\bf x})>0$ and assume the conditional probability $Pr( x_i'<x_i|{\bf x}_{<i})$
 with $x_i'$ being the random variable.
 
 The probability $p_F({\bf x})$ can be decomposed into a product 
\begin{equation}
 p_F({\bf x})=\prod_{i=1}^E p_F(x_i|{\bf x}_{<i})
\end{equation}
 
 The transformation from ${\bf x}$ to random number ${\bf z}\in [0,1]^E$ is defined as $F$
\begin{equation}
 z_i=F_i(x_i,{\bf x}_{<i})=\int_{-\infty}^{x_i} p_F(x_i'|{\bf x}_{<i}) dx_i'=Pr(x_i'\leq x_i|{\bf x}_{<i}).
\end{equation}
 Papamakarios et al\cite{PNRML21} showed that the inversion can be done element-by-element as
\begin{equation}
 x_i=(F_i(\cdot,{\bf x}_{<i}))^{-1}(z_i) \quad {\rm for} \quad i=1,2,\cdots,E.
\end{equation}
  Since $\frac{\partial F_i}{\partial x_j}=0$ for $i<j$ the determinant of Jacobian $J_F$ is a product of diagonal components
\begin{equation}
  det J_F({\bf x})=\prod _{i=1}^E\frac{\partial F_i}{\partial x_i}=\prod_{i=1}^E p_F(x_i|{\bf x}_{<i})=p_F({\bf x}).
\end{equation}
  The variable ${\bf z}$ is distributed uniformly in the open unit cube $(0,1)^E$.
  
  The transformation from $\bf u$ to $\bf z$ can be defined as
\begin{equation}
  z_i=G_i(u_i,{\bf u}_{<i})=\int_{-\infty}^{u_i} p_u(u_i'|{\bf u}_{<i})du_i'=Pr(u_i'\leq u_i|{\bf u}_{<i} ).
\end{equation}

We regard phonons as shock waves produced on Fermi surfaces and calculate plaquette actions derived from the fixed point action in $4D$\cite{DGHHN95} at $(2+1)D$ lattice points with the asymptotic high momentum action subtracted \cite{SF21b,SF21c,SFDS21d}. 

\begin{figure*}[htb]
\begin{minipage}{0.47\linewidth}
\begin{center}
\includegraphics[width=6.5cm,angle=0,clip]{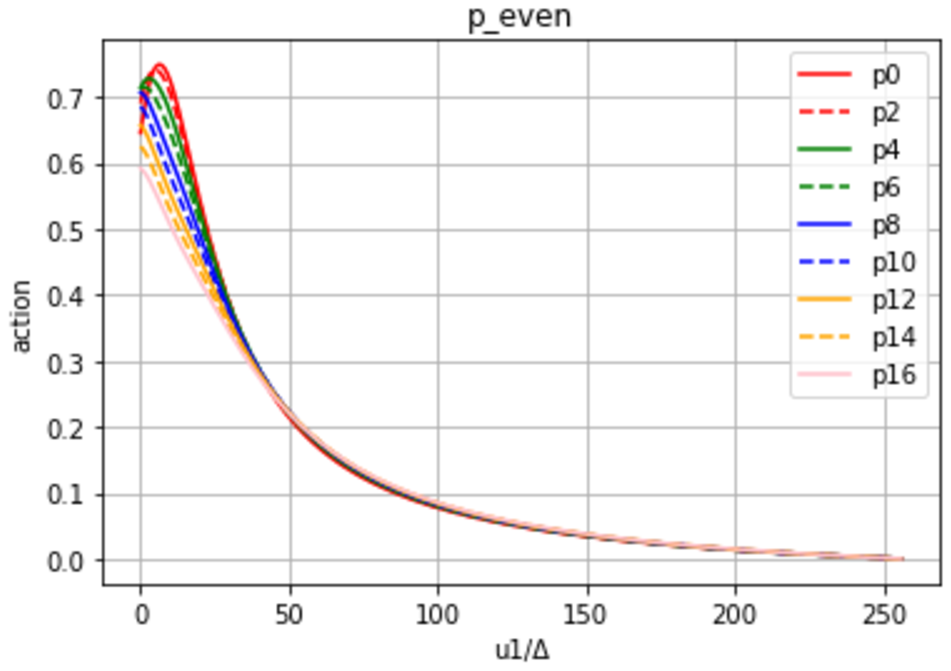}
\end{center}
\caption{ The plaquette action of $A$ type as a function of $u_1/\Delta$.  $u_2/\Delta=N_{even}$ $(0,2,\cdots,16)$.}
\end{minipage}
\hfill
\begin{minipage}{0.47\linewidth}
\begin{center}
\includegraphics[width=6.5cm,angle=0,clip]{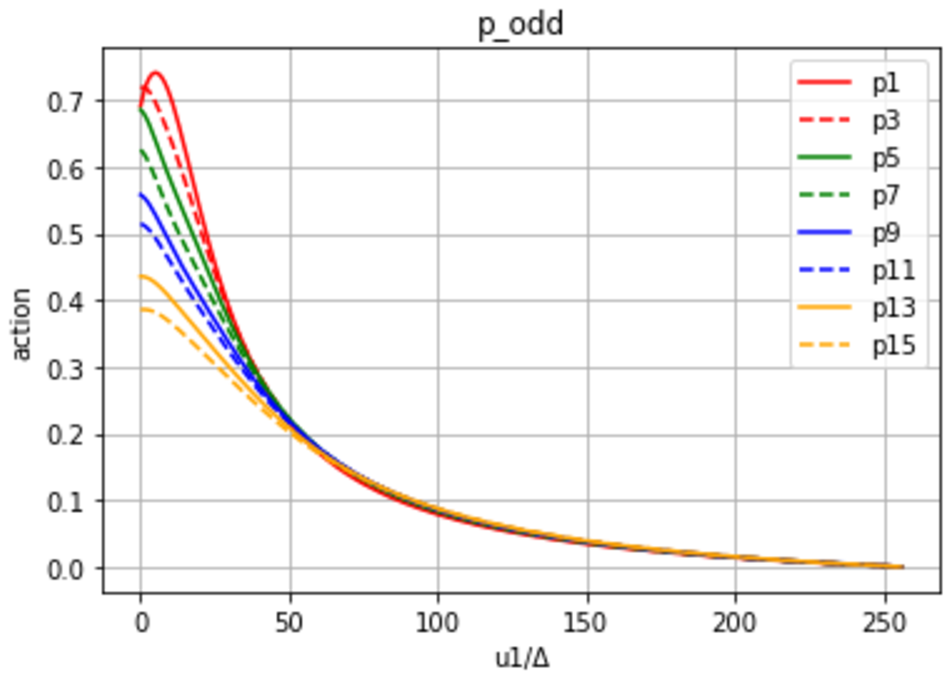}
\end{center}
\caption{ The plaquette action of $A$ type as a function of $u_1/\Delta$.  $u_2/\Delta=N_{odd}$ $(1,3,\cdots,15)$.}
\end{minipage}
\end{figure*} 
\begin{figure*}[htb]
\begin{minipage}{0.47\linewidth}
\begin{center}
\includegraphics[width=6.5cm,angle=0,clip]{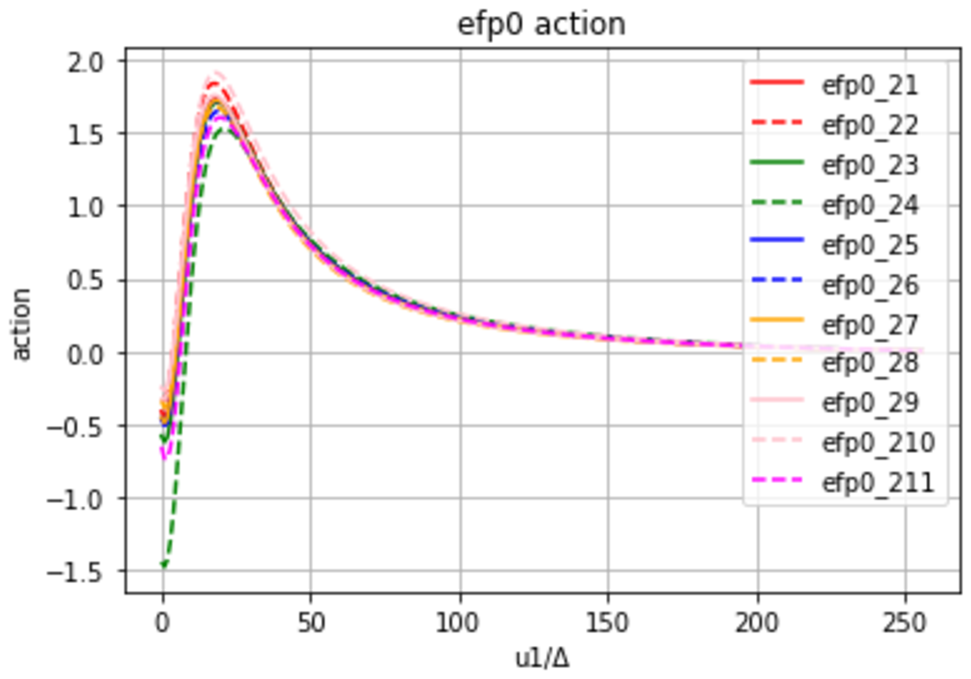}
\end{center}
\caption{ Sum of the plaquette action  and link action of $B$ type as a function of $u_1/\Delta$.  $u_2/\Delta=0$. The suffices $21,\cdots,211$ distinguish an order in the class 2 of random number sets used in the MC simulations.}
\end{minipage}
\hfill
\begin{minipage}{0.47\linewidth}
\begin{center}
\includegraphics[width=6.5cm,angle=0,clip]{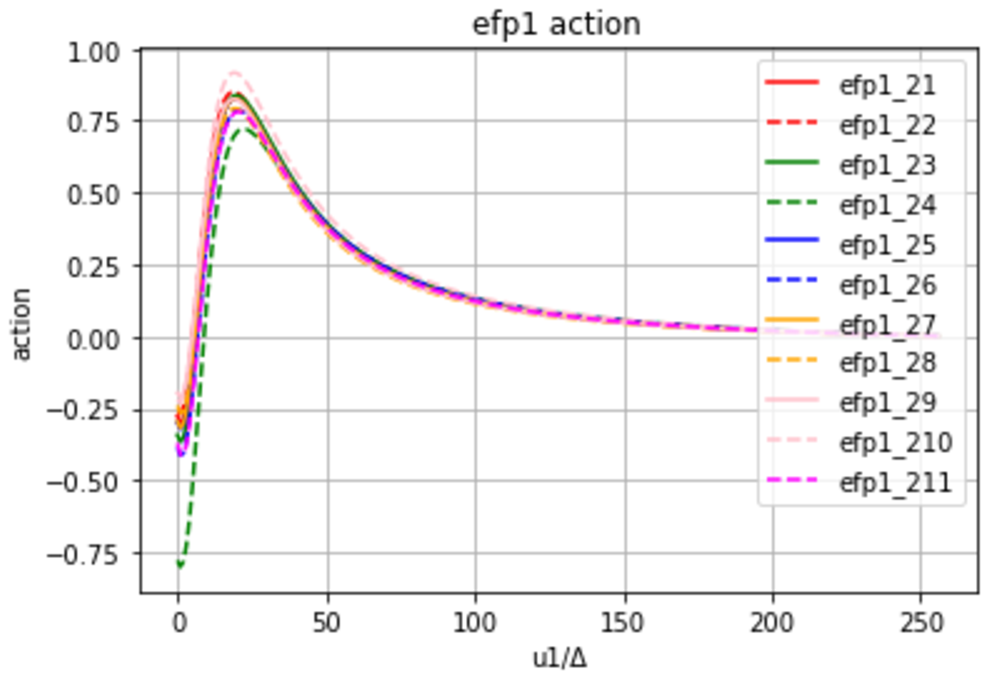}
\end{center}
\caption{ Sum of the plaquette action  and link action of $B$ type as a function of $u_1/\Delta$.  $u_2/\Delta=1$. The suffices $21,\cdots,211$ distinguish an order in the class 2 of random number sets used in the MC simulations.}
\end{minipage}
\end{figure*}

In Fig.1,2 the plaquette actiion of $A$ type loops as function of $u_1/\Delta$, where $\Delta$ is the lattice unit in momentum space at fixed $u_2/\Delta$ is even or odd are compared. Averages of Monte-Carlo (MC) simulations show that the slope of actions for $u_2/\Delta$ is odd is about twice of those for $u_2/\Delta$. 

Link actions of $B$ type loops for large momenta are asymptotically zero, but they depend on the clockwise rotating (f-link) or counterclockwise rotating. (e-link).  The relative strength of actions of e-link and f-link type depends on the random number sets used in MC. The average of e-link action and f-link action which are negative in the infrared region, plus plaquette action which are similar to that of $A$ type, obtained in a MC is shown in Fig.3,4. In the MC simulation, 50 orderings of 13 random numbers were considered as in the traveling salesman problem\cite{PM96}.

The maximum and minimum of the total action of $u_2/\Delta=0$ is about twice of those of $u_2/\Delta=1$. It reflects the difference of edge spin class of Kitaev\cite{FK11}, the former is real and the latter is quaternionic. 
This property reflects properties of CFT for $d=2,3$(mod 4) remarked by Hitzer(2022).
For a multivector $A_r\in{\mathcal G}_d$ of odd grade  $r=2s+1$ or even grade $r=2s$
\[
A_{2s+1}i_d=-i_d A_{2s+1}, \quad A_{2s}i_d=+i_d A_{2s}..
\] 
 We expect an oscilating pattern in the spectrum.
 
\section{Parameter search by machine learnings}
Recently, effectiveness of solving gauge theories using machine learnings or neural networks developed for Markov Chain Monte Carlo(MCMC) was realized.

Obtaining optimal weight function in normalizing flow of lattice MC using Python and PyTorch\cite{RLM22} was presented by several authors \cite{ABHKCRRS21}.

The Markov process consists of movement of $X_t, (t\geq 0=T)$ and probability $P_x, (x\in S)$ such that
$P_{t+s}(x,y)=\sum_z P_t(x,z)P_s(z,y) (t\in T, x,y\in S)$ is satisfied. To characterize the flow one defines a Hamiltonian $H({\bf \eta},{\bf v})=-log [p({\bf \eta},{\bf v})]$, phere $p({\bf \eta},{\bf v})$ is a joint distribution of the variable of interest $\bf \eta$ and $\bf v$. The hamiltonian flow is defined by 
\[
\frac{d({\bf \eta},{\bf v})}{dt}=(\frac{\partial H}{\partial {\bf v}}, -\frac{\partial H}{\partial {\bf\eta}}). 
\]

In a perceptron, which is a prototype of artificial neuron, one defines a decision function
$\sigma(z)$ where  $z=w_1 x_1+w_2 x_2+\cdots+w_m x_m$ is given by a weight vector ${\bf w}$ and  
input vector ${\bf x}$ as $z={\bf w}^T{\bf x}+b$.

The decision function is variant of unit step function
\[
\sigma(x)=\left\{\begin{array}{c}
                 1\quad {\rm if}\quad z\geq 0\\
                 0 \quad{\rm otherwise}
                 \end{array}\right.
\]

The weight function  $w_j$ and bias $b$ are defined sequentially and
\begin{eqnarray}
w_j &=&w_j+\Delta w_j \nonumber\\
b&=&b +\Delta b
\end{eqnarray}
where the output value $\hat y^{(i)}$  for an input $x^{(i)}$ is related by
\begin{eqnarray}
\Delta w_j &=&\eta(y^{(i)}-\hat y^{(i)})x_j^{(i)}\nonumber\\
\Delta b&=&\eta(y^{(i)}-\hat y^{(i)}).
\end{eqnarray}

One defines mean squared error (MSE) as
\begin{equation}
L({\bf w},b)=\frac{1}{2n}\sum_{i=1}^n (y^{(i)}-\sigma(z^{(i)}))^2,
\end{equation}
which is to be minimized.
 
In multi-layer neural network  (NN),  applied to recognitions of a hand-written number shows reduction of errors after training\cite{RLM22}. 

The algorithm consists of setting prior random number sets, forward setting and backward setting and calculation of Kullback-Leibler (KL) divergences. We want to extend the algorithm such that the information of f-link action is gained, learn optimum values of  e-link weight parameters. The difference of distributions are measured by R\'enyi entropy\cite{Fayngold13}.

\section{Perspective of Machine Learning}
In order to simulate propagation of phonetic waves in hysteretic media, we adopted a model that phonetic waves propagate in the sea of Weyl fermions.
To compare with experiments, we considered topological effects that emerge from fixed point actions on $(2+1)D$ lattices. We compared the Shannon entropy of several random number sets. We observed that a sample $\gamma$ among 4 samples, whose magnitude of the shift of e-link action from the average of e-link and f-link is close to that of f-link, 
the Shannon entropy becomes small.

The KL divergence does not show significant difference among random number sets.   

 In NDT, the random number set similar to $\gamma$ shows a strong signal, however R\'enyi entropy search will be helpful for more general situations.
 
 For optimization of the weight function in ML, CNN and RNN using stochastic optimization which is called adaptive moment estimation (ADAM) \cite{RLM22} would be appropriate. 
  In this method $|det J_\tau({\bf u};\phi)|$ is evaluated by introducing a scaling parameter $\alpha$, velocity parameters $\beta_1,\beta_2$ and gradients $g_1,\cdots , g_E$ at each epoch.
  
 At each epoch, CFT from $\bf u$ space to $\bf x$ space and inverse CFT from $\bf x$ space to $\bf u$ space is necessary.
In order to analyze the sum of plaquette actions, e-link actions and f-link actions, one-to-many RNN can be used. One-to-many reflects existence of many different hysteresis paths, which can be treated by Lie groupoids\cite{SFDS21d}.

The convolution of a Khokhlov-Zaboltskaya solitonic wave obtained by Lapidus and Rudenko\cite{LR92}, and its TR solitonic wave show presence of anomalous zero mode\cite{SFDS21a}, and detailed analysis of TR-NEWS convolution data will present hints on the gravitational anomaly suggested in \cite{RML12}.
Absence of zero modes may indicate mutual cancellations with the gravitational anomaly.

\begin{acknowledgements}
S.F. thanks RCNP of Osaka University for the support of using super computer SQUID and Prof. M. Arai for the spport of using a workstation in his laboratory. 
\end{acknowledgements}


\begin{thebibliography}{23}
\bibitem{Fink97} M. Fink, 
Phys. Today {\bf 50}, pp. 34-40, 1992.
\bibitem{RBCBF06} J-L. Robert, M. Burcher, C. Cohen-Bacrie and M. Fink, 
J. Acoust. Soc. Am.
{\bf 119}(6) pp.3848-3859 (2006).
\bibitem{GDSBMC08} T. Goursolle, S. Dos Santos, O. Bou Matar and S. Calle, 
Int. J..of Non-Linear Mechanics {\bf 43} pp.170-177, 2008. 
\bibitem{DSVP09} S. Dos Santos, S. Vejvodova and Z. Prevorovsky, 
J. Acoust. Soc. Am. {\bf 126} (5) (2009).
\bibitem{PM22} G. U. Patil and K. H. Matlack, 
Acta Mech {\bf 233} 1-46 (2022).
\bibitem{Haldane04} F.D.M. Haldane, 
Phys. Rev. Lett. {\bf 93}(20) 206602 (2004).

\bibitem{KM05} C.I. Kane and E.J. Mele, 
Phys. Rev. Letters {\bf 95}, 146802 (2005).


\bibitem{FK11} L. Fidkowski and A. Kitaev, 
Phys. Rev. {\bf B 83}, 075103 (2011).
\bibitem{RML12} S. Ryu, J.E. Moore and A.W.W. Ludwig,, 
Phys. Rev. {\bf B 85}, 045104 (2012). 


\bibitem{SFDS21a} S. Furui and S. DosSantos, 
arXiv:2010.09487 [physics.gen-ph] (v6) (2022).

\bibitem{DGHHN95} T. DeGrand, A. Hasenfratz, P. Hasenfratz and F. Niedermayer,  
Nucl. Phys.{\bf B454} 615-637 (1995), arXiv:9500031[hep-lat].

\bibitem{SF21b} S. Furui, 
arXiv:2105.06265[hep-lat] (2021).

\bibitem{SF21c} S. Furui, 
arXiv: 2107.07880[hep-lat] (2021).

\bibitem{SFDS21d} S. Furui and S. Dos Santos, 
arXiv: 2112.13648 (v2)[hep-lat] (2021).
\bibitem{PM96} A.G. Percus and O.C. Martin, 
Phys. Rev. Lett. {\bf 76}(8) 1188-1191(1996).
\bibitem{Fayngold13} M. Fayngold and V. Fayngold, {\it Quantum Mechanics and Quantum Information}, Wiley-VCH (2013).
\bibitem{RLM22} S. Raschka, Y. Liu and V. Mirjalili, {\it Machine Learning with Pytorch and Scikit-Learn}, Packt Pub. Birmingham-Munbai (2022).
 
\bibitem{ABHKCRRS21} M.S. Albergo et al. 
arXiv: 2101.08176v3[hep-lat] (2021).
\bibitem{PNRML21} G. Papamakarios et al.  
J. of Machine Learning Research {\bf 22} 1-64 (2021).
\bibitem{Ye22} J.C. Ye, ``Geometry of Deep Learning'',Springer (2022).
\bibitem{KABCH20} G. Kanwar et al 
Phys. Rev. Lett. {\bf 125}, 121601 (2020).


\bibitem{Hitzer22} Eckhard Hitzer, {\it Quaternion and Clifford Fourier Transforms}, CRC Press, Boca Raton, London, New York (2022),


\bibitem{LR92}
Yu.R. Lapidus and O.V. Rudenko, 
Sov. Phys. Acoust. {\bf 38} (2) (1992).
\end{thebibliography}
\end{document}